\documentclass[utf8]{FrontiersinHarvard} 

\usepackage{url, hyperref, lineno, microtype} 
\usepackage[onehalfspacing]{setspace}
\usepackage{doi}
\usepackage{soul}

\def\keyFont{\fontsize{8}{11}\helveticabold }
\def\firstAuthorLast{Gieseler {et~al.}} 
\def\Authors{
Jan Gieseler\,$^{1,*}$, 
Nina Dresing\,$^{1}$, 
Christian Palmroos\,$^{1}$, 
Johan L. Freiherr von Forstner\,$^{2,3}$,
Daniel J. Price\,$^{4}$,
Rami Vainio\,$^{1}$, 
Athanasios Kouloumvakos\,$^{5}$,
Laura Rodr\'iguez-Garc\'ia\,$^{6}$,
Domenico Trotta\,$^{7}$,
Vincent G\'enot\,$^{8}$,
Arnaud Masson\,$^{9}$,
Markus Roth\,$^{10}$,
Astrid Veronig\,$^{11}$
}

\def\Address{
$^{1}$Space Research Laboratory, Department of Physics and Astronomy, University of Turku, Turku, Finland\\
$^{2}$Institute of Experimental and Applied Physics, Kiel University, Kiel, Germany\\
$^{3}$Now at Paradox Cat GmbH, München, Germany\\
$^{4}$Department of Physics, University of Helsinki, Helsinki, Finland\\
$^{5}$Applied Physics Laboratory, The Johns Hopkins University, Laurel, MD, USA\\
$^{6}$Universidad de Alcalá, Space Research Group, Alcalá de Henares, Madrid, Spain \\
$^{7}$ The Blackett Laboratory, Department of Physics, Imperial College London, London UK \\
$^{8}$Institut de Recherche en Astrophysique et Planétologie, CNRS, CNES, Université Paul Sabatier, Toulouse, France\\
$^{9}$ESAC Science Data Centre, European Space Agency, Madrid, Spain\\
$^{10}$Th\"uringer Landessternwarte, Tautenburg, Germany\\
$^{11}$Institute of Physics \& Kanzelh\"ohe Observatory for Solar and Atmospheric Research, University of Graz, Austria
}
\def\corrAuthor{Jan Gieseler, University of Turku, 20500 Turku, Finland}

\def\corrEmail{jan.gieseler@utu.fi}

\begin{document}
\onecolumn
\firstpage{1}

\title[Solar-MACH]{Solar-MACH: An open-source tool to analyze solar magnetic connection configurations} 

\author[\firstAuthorLast ]{\Authors} 
\address{} 
\correspondance{} 

\extraAuth{}

\maketitle
\begin{abstract}
The Solar MAgnetic Connection HAUS\footnote{\href{https://www.cosmos.esa.int/web/esdc/archives-user-groups/heliophysics}{Heliophysics Archives USer group at ESA}} tool (Solar-MACH) is an open-source tool completely written in Python that derives and visualizes the spatial configuration and solar magnetic connection of different observers (i.e., spacecraft or planets) in the heliosphere at different times. For doing this, the magnetic connection in the interplanetary space is obtained by the classic Parker Heliospheric Magnetic Field (HMF). In close vicinity of the Sun, a Potential Field Source Surface (PFSS) model can be applied to connect the HMF to the solar photosphere. Solar-MACH is especially aimed at providing publication-ready figures for the analyses of Solar Energetic Particle events (SEPs) or solar transients such as Coronal Mass Ejections (CMEs). It is provided as an installable Python package (listed on PyPI and conda-forge), but also as a web tool at \href{https://solar-mach.github.io}{solar-mach.github.io} that completely runs in any web browser and requires neither Python knowledge nor installation. The development of Solar-MACH is open to everyone and takes place on GitHub, where the source code is publicly available under the BSD 3-Clause License. Established Python libraries like \texttt{sunpy} and \texttt{pfsspy} are utilized to obtain functionalities when possible. In this article, the Python code of Solar-MACH is explained, and its functionality is demonstrated using real science examples. In addition, we introduce the overarching SERPENTINE project, the umbrella under which the recent development took place.



\tiny
 \keyFont{ \section{Keywords:} Python, Software Package, Solar Energetic Particle (SEP), Corona, PFSS, Coronal Mass Ejection (CME), Spacecraft, Heliosphere} 
\end{abstract}

\section{Introduction}



The \emph{Solar energetic particle analysis platform for the inner heliosphere} (\href{https://serpentine-h2020.eu/}{SERPENTINE}, 2021--2024) is a 42-month long project funded through the H2020-SPACE-2020 call of the European Union's \href{http://ec.europa.eu/programmes/horizon2020}{Horizon 2020 framework programme}. The project addresses several outstanding questions on the origin of solar energetic particle (SEP) events and provides an advanced data analysis and visualization platform that will benefit the whole heliophysics community. It utilizes the most recent European and US missions, i.e., Solar Orbiter \citep{Muller2020}, Parker Solar Probe \citep{Fox2016} and BepiColombo \citep{Benkhoff2021}. These observations are complemented with supporting data from several current missions near Earth’s orbit as well as ground-based radio imaging and spectroscopic observations by the European Low Frequency Array \citep[LOFAR;][]{vanHaarlem2013}.

SEP events are large and sporadic outbursts of charged particle radiation from the solar corona that are related to solar eruptions such as flares and coronal mass ejections \citep[CMEs; e.g.,][]{Reames1999}. They can be classified as impulsive and gradual events, based on their duration and the duration of the related solar X-ray flare. Impulsive SEP events are associated with impulsive flares and narrow CMEs, and they are enriched in electrons, $^3$He isotope and heavy ions. Gradual SEP events are associated with gradual solar X-ray flares and fast and wide CMEs, and their abundances resemble nominal coronal abundances \citep{Reames2013, Desai2016}. Gradual events are usually broader in their helio-longitudinal extent than impulsive events and their intensities are also typically larger, making them the main concern of spacecraft operations and crews \citep{Vainio2009}.

The primary reason for the broad spatial extent of some gradual events is still unresolved. It could be due to a broad source, like a global coronal shock driven by a CME \citep[e.g.,][]{Lario2016}, due to efficient particle transport processes across the heliospheric magnetic field, or as a result of both mechanisms \citep[e.g.,][]{Dresing2012, Rodriguez-Garcia2021}. The main objective of SERPENTINE is to pinpoint the primary causes of large gradual and widespread SEP events. To address this objective, SERPENTINE will answer the following open science questions:
\begin{itemize}
    \item [Q1:] What are the primary causes for widespread SEP events observed in the heliosphere?

    \item [Q2:] What are the shock acceleration mechanisms responsible for accelerating ions from thermal/suprathermal energies to near-relativistic energies in the corona and in the interplanetary medium?

    \item [Q3:] What is the role of shocks in electron acceleration in large gradual and widespread events? How does it relate to ion acceleration and what is its importance relative to flare acceleration?
    
\end{itemize}
To reach these goals and also to broaden the impact of the project, SERPENTINE will develop and release to the community a platform of tools for analyzing SEP events. Furthermore, the tools may also be useful for the broader heliospheric community looking at different aspects of solar activity or solar wind phenomena. Part of this platform will be a JupyterHub server that provides free access to the tools developed by the SERPENTINE project as Jupyter Notebooks, without requiring any installations beyond a web browser.
This manuscript and the accompanying papers \citep{PyThea2022, Palmroos2022, Trotta2022} will present the first batch of these tools to the heliophysics community.  

\begin{figure}[ht!]
\begin{center}
\includegraphics[width=\textwidth]{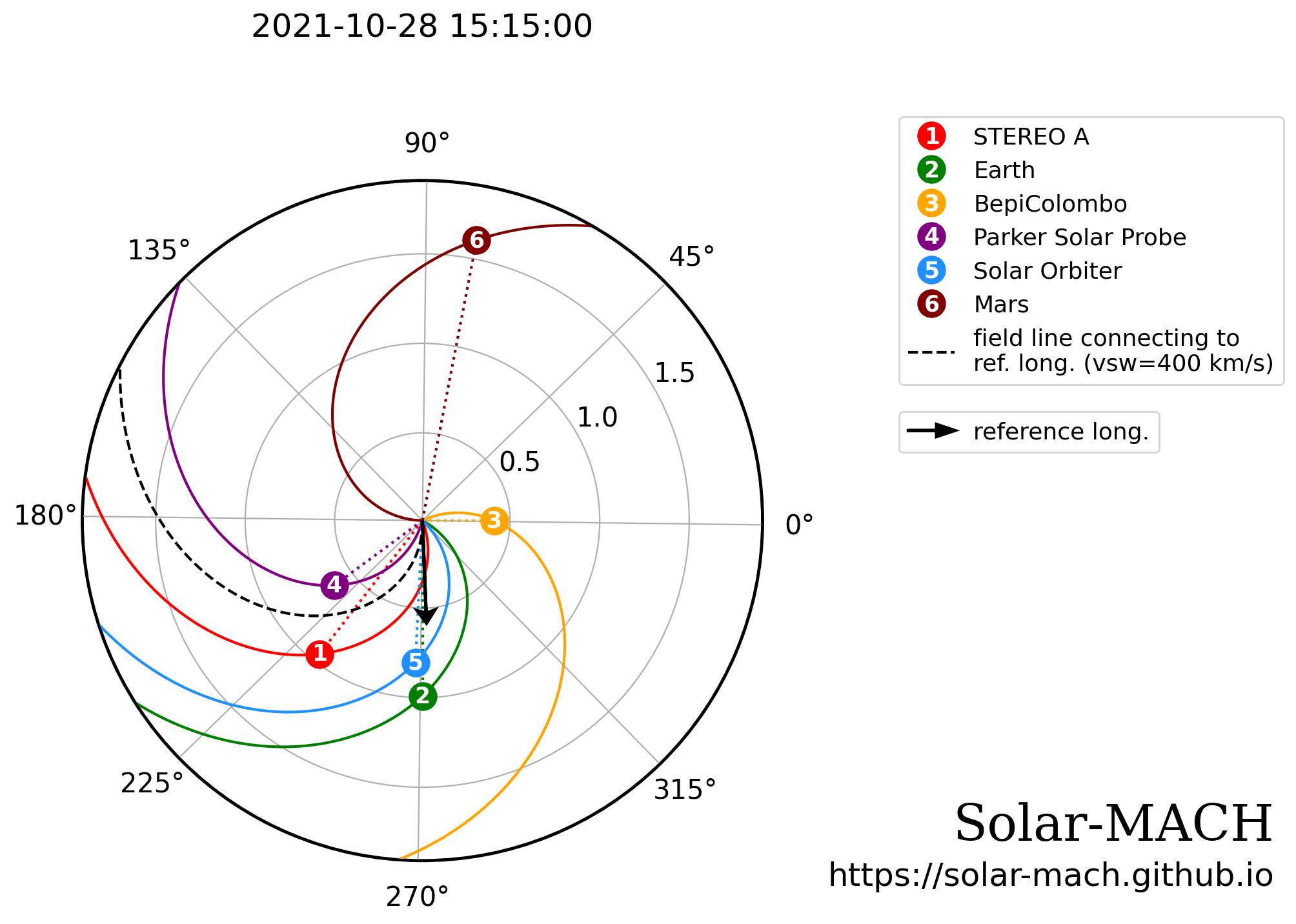}
\end{center}
\caption{
Solar-MACH plot for the time of the ground-level enhancement (GLE) event on 28 October 2021. Numbered symbols indicate the observers' locations and the spiral lines corresponding HMF lines connecting them to the Sun. Radial distance is provided in astronomical units (AU), and the angular information is given in Carrington longitude. The arrow points out the freely choosable location of a ``reference" (e.g., a solar flare) at the Sun, and the dashed spiral line indicates a corresponding HMF line originating at that position.
}
\label{fig:solarmach}
\end{figure}

Because SEPs are measured in situ as enhancements of the energetic particle fluxes, the presence of multiple, well-separated observers is indispensable to study widespread SEP events \citep[e.g., ][]{Dresing2014, Richardson2014}. The new space missions in combination with established spacecraft form a fleet that is ideal for this purpose, as it covers varying heliocentric distances and large longitudinal ranges around the Sun. The different orbits of the multiple spacecraft, in combination with varying source regions at the Sun, constantly form new constellations, building the baseline for in-depth SEP event analyses. The first released tool of SERPENTINE, Solar-MACH, provides the user with a quick overview of these specific constellations for a selected time, as shown in the example of Fig.~\ref{fig:solarmach}. ``Solar-MACH" is an abbreviation for \textit{Solar MAgnetic Connection HAUS}, with HAUS standing for the \href{https://www.cosmos.esa.int/web/esdc/archives-user-groups/heliophysics}{\textit{Heliophysics Archives USer} group at ESA}. ``MACH" is intended to be pronounced like the \textit{Mach} number.

\section{Method}
The main functionality of Solar-MACH is to provide the user with a polar plot showing a Sun-centered top view of the heliographic equatorial plane with the constellation of different observers for a given time, as shown in Fig.~\ref{fig:solarmach}. Additional information can be added, such as indicating the magnetic connection of each observer to the Sun as given by an idealized classic Parker heliospheric magnetic field line, whose curvature is dependent on the radial solar wind speed. This and other options are described in Sect.~\ref{sect:general}.

All software operations are performed within Python 3 \citep[e.g.,][]{VanRossum2009}; the corresponding package \texttt{solarmach} is described in Sect.~\ref{sect:python_package}. In addition, a web tool is provided that 
requires neither Python knowledge nor installation. This web tool and its technical background are presented in Sect.~\ref{sect:streamlit}.

\subsection{General functionality}
\label{sect:general}
\subsubsection{Ephemeris}

After the user provides a specific date and time, the spatial coordinates of all requested observers are obtained dynamically from \href{https://ssd.jpl.nasa.gov/horizons/}{JPL Horizons}, an online solar system data and ephemeris computation service that is maintained by the Solar System Dynamics Group of the Jet Propulsion Laboratory (JPL). In addition to the web interface, JPL Horizons provides an API for programmatic control. This API is utilized through version 4.0.5 \citep{Mumford2022} of the \texttt{sunpy} open-source software package \citep{Sunpy2020}, which itself uses Astropy's \citep{Astropy2013, Astropy2018, Astropy2022} affiliated package \href{https://www.astropy.org/astroquery/}{\texttt{astroquery}} \citep{Ginsburg2021} for this functionality. The ephemeris information is returned per observer within Python as a \href{https://docs.astropy.org/en/stable/api/astropy.coordinates.SkyCoord.html}{\texttt{SkyCoord}} object, which inherently provides coordinate system transformations. This allows easily supporting different heliographic coordinate systems. At the moment, the user can decide whether all operations should be carried out in Stonyhurst or Carrington coordinates \citep[e.g.,][]{Thompson2006}, with the latter being the default setting.

\subsubsection{Magnetic connectivity}
One important property for analyzing SEP events is the magnetic connection of each observer to the Sun. A simple approximation for this connection can be obtained by assuming an ideal Archimedean spiral configuration for the heliospheric magnetic field \citep{Parker1958}, which is defined in spherical coordinates by:
\begin{eqnarray}
\varphi(r) = \varphi_0 + \frac{\omega(\vartheta)}{v_{sw}}\cdot(R-r)\cdot \cos{\vartheta}, \label{eq:parkerspiral}
\end{eqnarray}
with heliographic longitude $\varphi$ as a function of radial distance $r$, heliographic latitude $\vartheta$, differential solar rotation frequency $\omega(\vartheta)$, radial solar wind speed $v_{sw}$, as well as radial distance $R$ and longitude $\varphi_0$ of the observer. 
The differential solar rotation frequency $\omega(\vartheta)$ takes into account that the Sun's rotation speed varies with latitude. Here, we apply an empirical model by \citet{PoljancicBeljan2017} that describes $\omega(\vartheta)$ by $14.50 - 2.87 \sin^2{\vartheta}$ (deg/day), which corresponds to $14.50$ (deg/day) $= 1.678241\cdot 10^{-4}$ (deg/s) at the heliographic equator. For the two-dimensional visualization, all spherical coordinates are projected to the heliographic equatorial plane.
The resulting field lines are shown in Fig.~\ref{fig:solarmach} as solid spiral lines, with color-coding of the respective spacecraft.
As the spatial coordinates are dynamically obtained for each spacecraft, the solar wind speed $v_{sw}$ is the only free parameter that needs to be provided for each observer in units of $\textrm{km}/\textrm{s}$. By default, this is assumed to be $400\ \textrm{km}/\textrm{s}$. Automatically obtaining real measurements of the solar wind speed is planned as an option for the future.
Applying Eq.~\ref{eq:parkerspiral} for $r=R_\odot$ provides the magnetic footpoint coordinate for each spacecraft, that is, the heliographic longitude where the assumed magnetic field line originates at the Sun. This information is especially important in order to connect in situ measurements of SEPs with remote-sensing observations of the associated source regions at the Sun. We must point out, though, that this analysis is just an estimation that may deviate significantly from reality. Employing this tool for magnetic backmapping purposes comes with the known caveats of the simple assumptions of ballistic backmapping, which are a constant and totally radial solar wind speed as well as the absence of interplanetary structures (e.g., interplanetary CMEs) that could deform the ideal Parker field.

\subsubsection{Reference location}
\label{sect:reference}
To further assist this analysis, the location of a so-called ``reference" at the Sun (and a corresponding $v_{sw}$) can be provided to the tool, for example, the position of a flare that is assumed to be related with the SEP event. This information is then used on the one hand to visually indicate the position of the reference (i.e., flare) in the plot with a black arrow (cf. Fig.~\ref{fig:solarmach}) and its corresponding heliospheric magnetic field line, which helps to quickly estimate which spacecraft are possible observers of particles originating at this flare. On the other hand, the separation angles between the position of the reference and the magnetic footpoints of the different spacecraft are calculated. 
This information can then be obtained for further use in a table that also contains all other coordinate parameters provided by JPL Horizons, as demonstrated in the bottom of Fig.~\ref{fig:notebook}.

\subsubsection{Plotting options}
\label{sect:general_plotting}
In addition to the functional options described so far, Solar-MACH provides some plotting settings that enable users to have several options of visualization customization:
\begin{itemize}
    \item Deactivate plotting of the heliospheric magnetic field lines.
    \item Add a straight line from the Sun to each observer, indicating the line of sight.
    \item Use numbered markers for the spacecraft to help with color vision deficiency issues or if the plot needs to be converted to grayscale.
    \item Provide a longitudinal offset for the polar plot to define where the Earth is located (by default $270^\circ$, i.e., at \textit{six o'clock}).
    \item Make the background of the plot transparent.
\end{itemize}

\subsection{Python package \textit{solarmach}}
\label{sect:python_package}
\begin{figure}[hp!]
\begin{center}
\includegraphics[height=0.94\textheight]{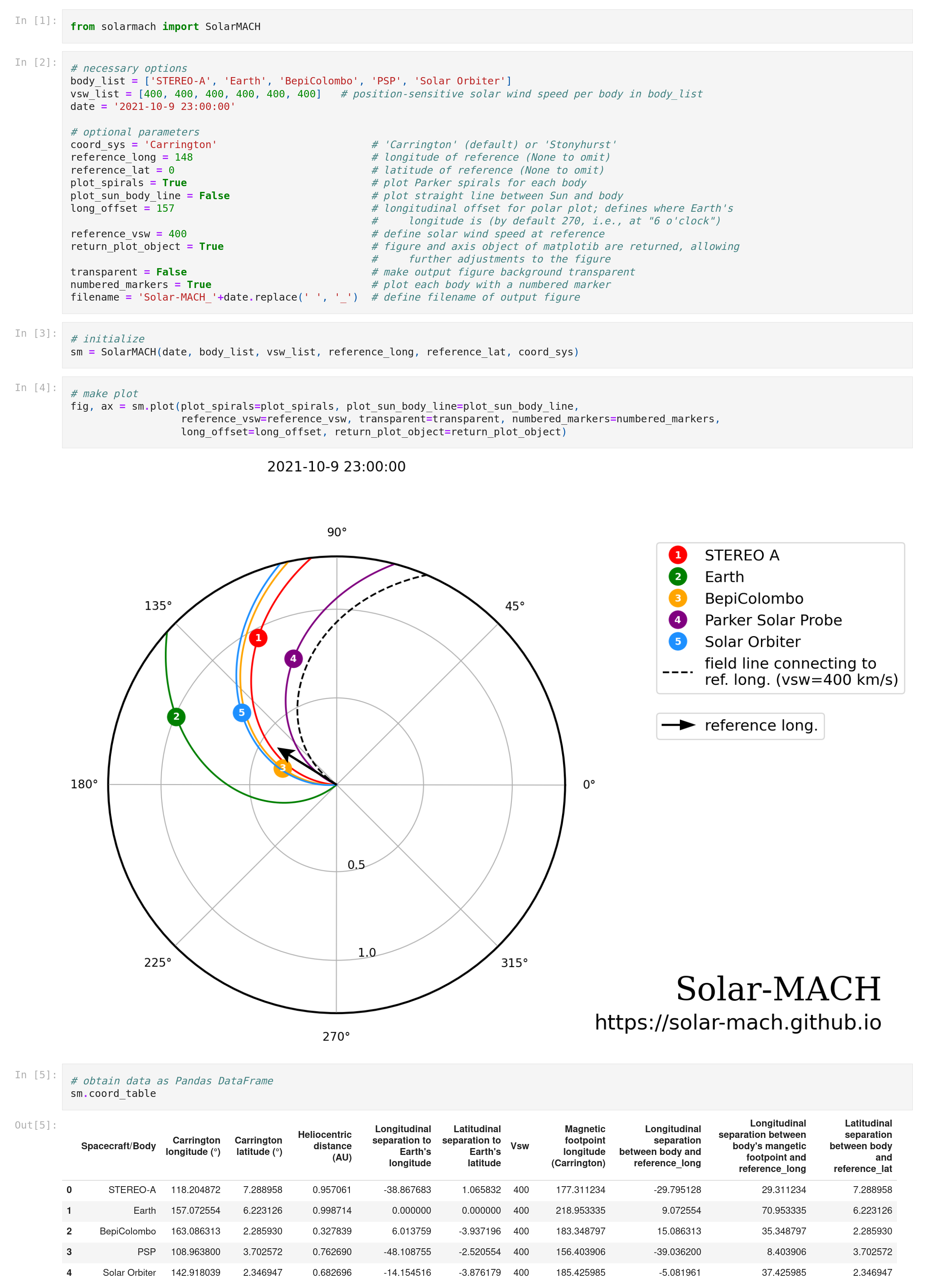}
\end{center}
\caption{Example of using the Python package \texttt{solarmach} in a Jupyter Notebook for the SEP event on 9 October 2021}\label{fig:notebook}
\end{figure}
The Solar-MACH Python code is made available as an installable Python package called \texttt{solarmach} \citep{solarmach2022} that is listed on \href{https://pypi.org/project/solarmach/}{PyPI} and \href{https://anaconda.org/conda-forge/solarmach/}{conda-forge}, i.e., it can be installed using either the \texttt{pip} or the \texttt{conda} command line tool, which are by far the most widely used means of installing Python packages. Installing the package in this way ensures that all other required packages are also installed.

An example of the standard workflow for the package is shown in a Jupyter Notebook in Fig.~\ref{fig:notebook}. This specific Notebook is also available from the subfolder \texttt{examples} in the corresponding \href{https://github.com/jgieseler/solarmach/}{GitHub repository}. The workflow consists of the following steps (each step is represented by one code cell in the example Notebook in Fig.~\ref{fig:notebook}):
\begin{enumerate}
    \item Import the \texttt{SolarMACH} class
    \item Provide necessary options and optional parameters
    \item Initialize the \texttt{SolarMACH} object with these settings
    \item Generate a plot from the \texttt{SolarMACH} object (optionally)
    \item Obtain the data as a Pandas DataFrame from the \texttt{SolarMACH} object (optionally)
\end{enumerate}
The Pandas DataFrame obtained in the last step does not only display the dataset in a tabular form, but also provides the user with a ready-to-use dataset for further analyses, ultimately enabling fast diagnostic pipelines \citep{Mckinney2010}. Furthermore, through the optional settings in step 2, it is possible to either provide a file name and ending under which the plot will then be saved (where the file type is automatically determined, e.g., ``file.png" or ``file.pdf"), or to return the matplotlib \citep{Hunter2007} plot object in order to further manipulate it. 

A Jupyter Notebook with more detailed descriptions and various use cases is available \href{https://github.com/serpentine-h2020/serpentine/tree/main/notebooks/solarmach}{in the SERPENTINE software repository} \citep{SERPENTINE2022}. This Notebook contains examples on how to connect the Solar-MACH outputs with other Python functionalities. One example is the creation of an animated GIF file that shows a time-lapse of the spacecraft constellation over a given time period (see \href{https://raw.githubusercontent.com/serpentine-h2020/serpentine/main/notebooks/solarmach/solarmach.gif}{here} for an example). In addition, this extended Notebook provides the functionality to continue the interplanetary magnetic connections to the solar corona with a PFSS model, which is described in more detail in Sect.~\ref{sect:pfss}.

\subsection{Further backmapping with PFSS}
\label{sect:pfss}
\begin{figure}[ht!]
\begin{center}
\includegraphics[width=\textwidth]{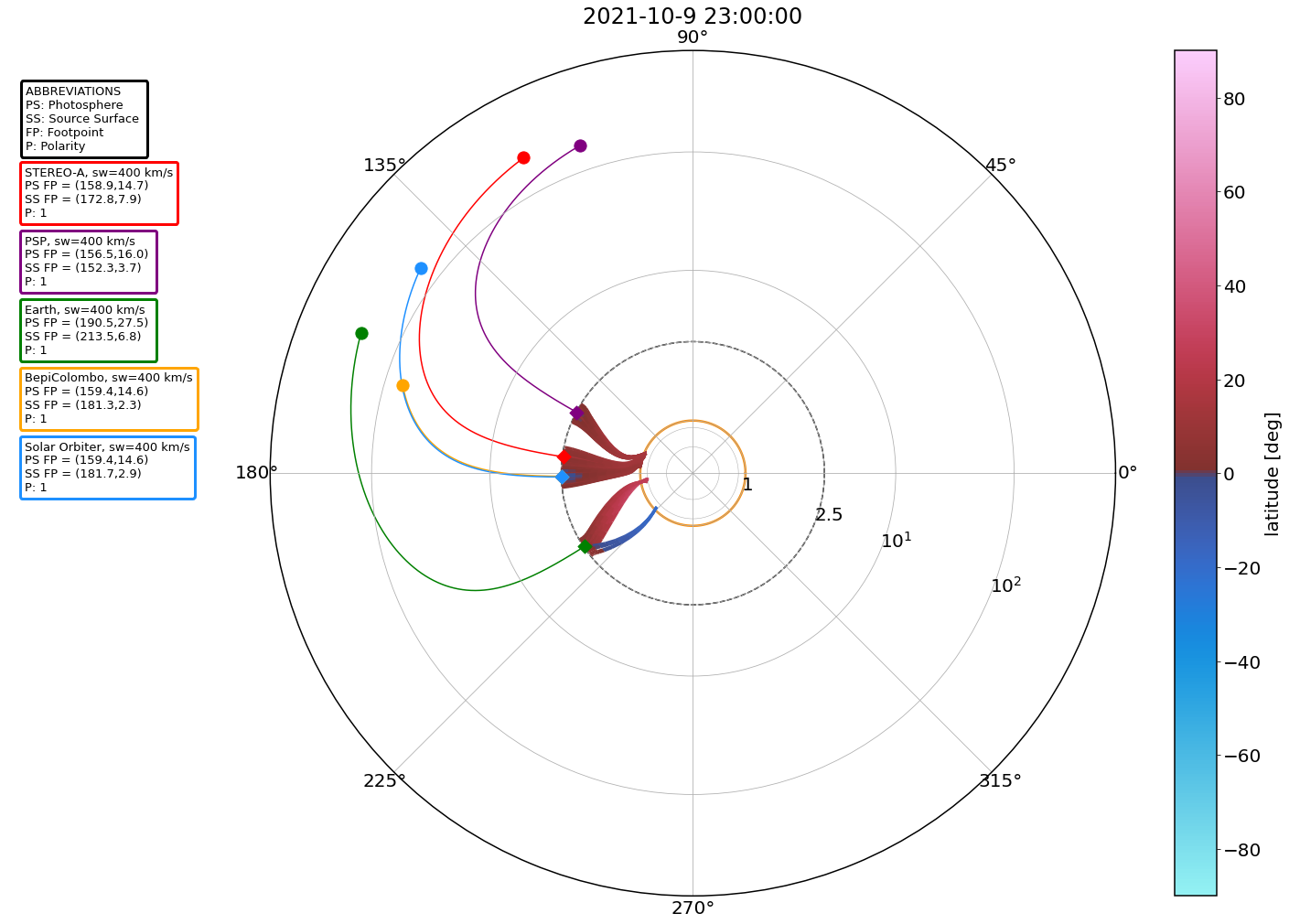}
\end{center}
\caption{PFSS backmapping output of Solar-MACH for the SEP event on 9 October 2021. Note the different scaling of the radial axis compared to Fig.~\ref{fig:notebook} (see text for details). The yellow circle represents the photosphere at one solar radius.}\label{fig:pfss}
\end{figure}

The magnetic field configuration close to the Sun in the corona differs drastically from the heliospheric magnetic field (HMF) and cannot be described using the simple Parker spiral approach given by Eq.~\ref{eq:parkerspiral}. In order to further extend the magnetic connectivity from an observer towards the solar surface, a Potential Field Source Surface (PFSS) model \citep[e.g.,][and references therein]{Mackay2012} is applied by using version 1.1.2 \citep{Stansby2022} of the \texttt{pfsspy} open-source software package \citep{Stansby2020}. 
The main purpose is to calculate a set of open magnetic field lines in a defined circular area around the point where an idealized HMF line connects an interplanetary observer to the source surface, at which the field lines are forced by the PFSS model to be radial. This is illustrated in Fig.~\ref{fig:pfss} for the situation on 9 October 2021. The outer colored field lines correspond to those of Fig.~\ref{fig:notebook}, only that the radial distance is now plotted in a semilogarithmic scale, so that the situation close to the Sun can be depicted in more detail. The radial axis is given in units of solar radii, with a linear scale up to where the source surface is assumed (this is a free parameter, by default set to 2.5 solar radii). Further out, a logarithmic scale is used, which explains the slightly different shape of the HMF lines compared to Fig.~\ref{fig:notebook}. Below the source surface, for each HMF line a set of open field lines is calculated and traced back to the photosphere. Color-coding gives additional information about the corresponding latitudes, as depicted by the color bar on the right. Figure~\ref{fig:pfss23} presents alternative ways to illustrate the results, where the set of coronal field lines obtained for the HMF line connecting to the Earth is shown in a two-dimensional side view (Fig.~\ref{fig:pfss23}a) or in a freely rotatable three-dimensional view (Fig.~\ref{fig:pfss23}b). These figures illustrate a situation in which the magnetic connection below the source surface is ambiguous as the bundle of field lines splits, each leading to a different region at the photosphere. Although the tool cannot provide an unambiguous backmapped magnetic footpoint location, such a result contains important information for the user, which are valuable for further analyses.
\setcounter{subfigure}{0}
\begin{subfigure}
\setcounter{subfigure}{0}
    \centering
    \begin{minipage}[b]{0.45\textwidth}
        \includegraphics[width=\linewidth]{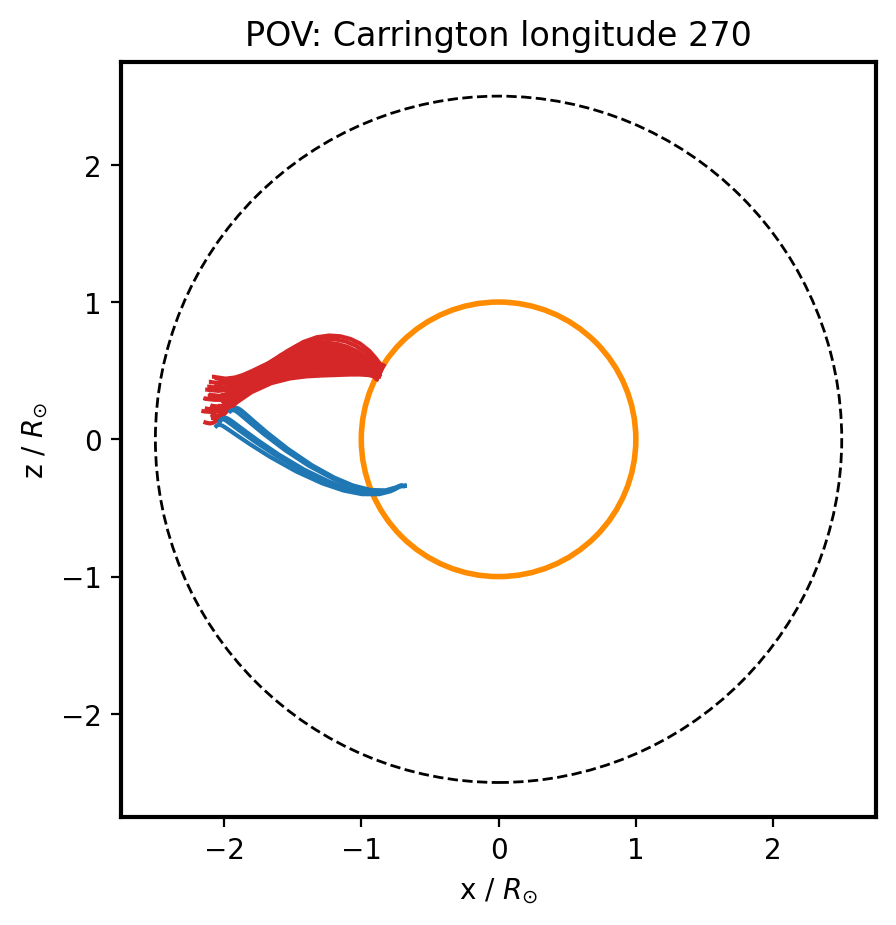}
        \label{fig:pfss2}
    \end{minipage}  
\setcounter{subfigure}{1}
    \begin{minipage}[b]{0.45\textwidth}
        \includegraphics[width=\linewidth]{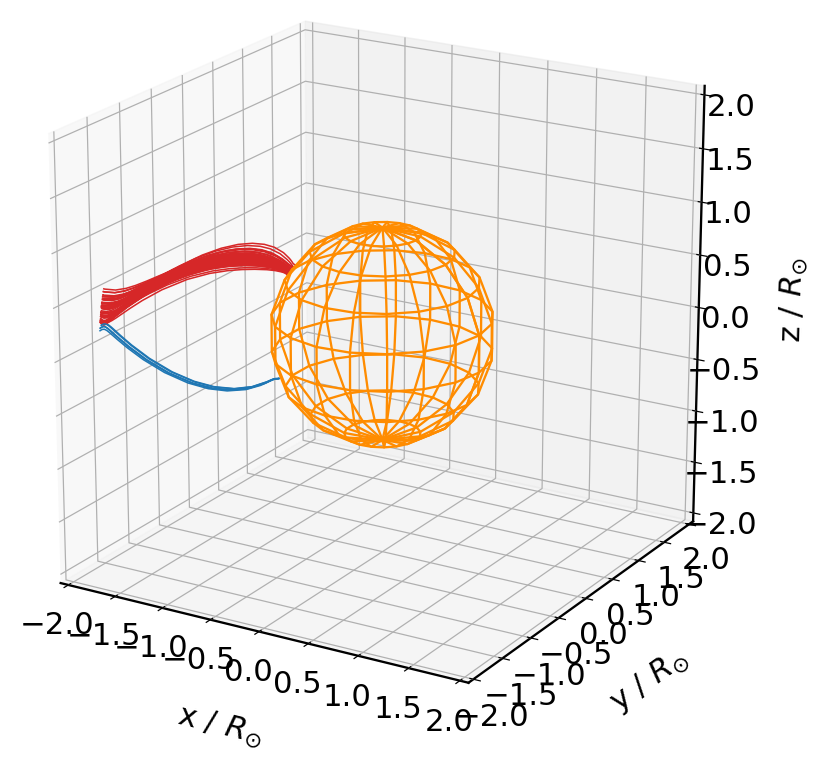}
        \label{fig:pfss3}
    \end{minipage}
\setcounter{subfigure}{-1}
    \caption{Detailed view of Earth's magnetic connection from the PFSS backmapping example shown in Fig.~\ref{fig:pfss}. \textbf{(A)} Two-dimensional side view with the point of view (POV) at Carrington longitude 270$^\circ$ (i.e., the observer is located at \textit{six o'clock} in Fig.~\ref{fig:pfss}), \textbf{(B)} freely rotatable three-dimensional view. The two colors of the field lines indicate the corresponding magnetic polarity.}
    \label{fig:pfss23}
\end{subfigure}

At the moment, the PFSS functionality of Solar-MACH is only provided through the Jupyter Notebook \href{https://github.com/serpentine-h2020/serpentine/tree/main/notebooks/solarmach}{in the SERPENTINE software repository} \citep{SERPENTINE2022} that has already been mentioned in Sect.~\ref{sect:python_package}. This part has not yet been added to the \texttt{solarmach} package, but this is planned for the near future, subsequently making it available through the Streamlit web tool (cf. Sect.~\ref{sect:streamlit}), too.
In order to be able to obtain Helioseismic and Magnetic Imager (HMI) synoptic maps through the Joint Science Operations Center (JSOC) at Stanford University, which are needed as input for the PFSS model in the current version, the user needs to register once at \url{http://jsoc.stanford.edu/ajax/register_email.html}. For the future, especially for the integration into the Streamlit web tool, we aim at removing this requirement.

\subsection{Streamlit web tool}
\label{sect:streamlit}
\begin{figure}[hp!]
\begin{center}
\includegraphics[height=0.95\textheight]{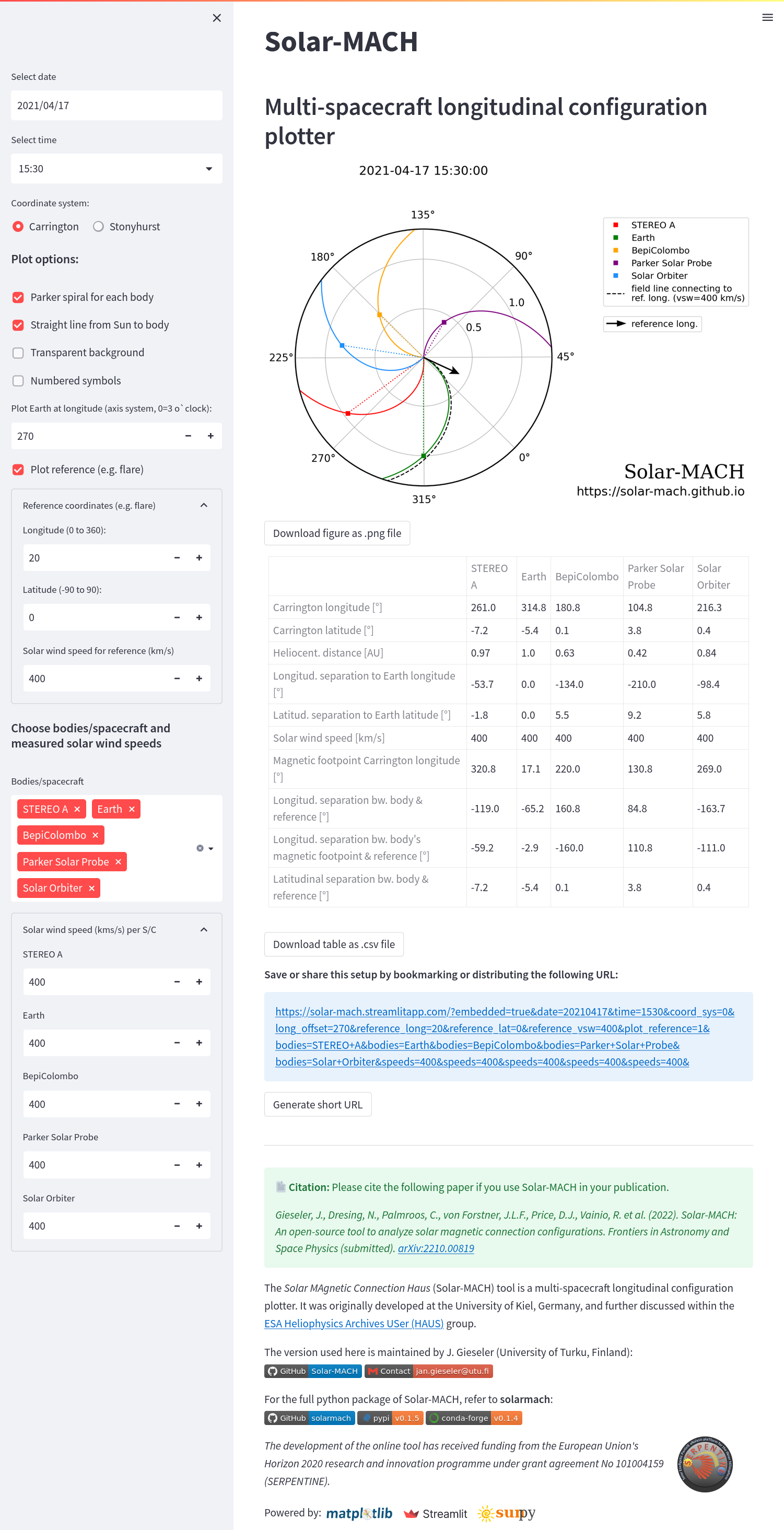}
\end{center}
\caption{The Streamlit web version at \href{https://solar-mach.github.io}{solar-mach.github.io}}\label{fig:streamlit}
\end{figure}

While all the functionality of Solar-MACH depends fully on Python, a web tool is provided online at \href{https://solar-mach.github.io}{solar-mach.github.io} that completely runs in the web browser, features a fully graphical user interface (GUI), and requires neither Python knowledge nor installation. It is implemented using the open-source Python package \href{https://github.com/streamlit/streamlit/}{\texttt{streamlit}}. This package provides an easy way to set up a GUI to Python functions that is made available through a web server. 
It is possible to run it locally on a computer (if Python and all other requirements are installed), offering a full GUI to the user instead of running the code in a Jupyter Notebook or terminal. But the main purpose is to deploy the Streamlit application to a web hosting service, because only then it provides all the underlying Python functionalities without the need to install any kind of (Python) software. A screenshot of the web tool is shown in Fig.~\ref{fig:streamlit}. We emphasize that the possibility to use the web tool is invaluable for getting started quickly in a non-code environment, which is especially important for new users.

In addition to quickly obtaining an overview of the observer constellation when a SEP event just took place, another important functionality of the web tool is to bookmark or share a given combination of options that has been set through the GUI. After all settings have been done, a specific URL for this configuration is shown at the bottom of the web page (blue box in Fig.~\ref{fig:streamlit}). An option is provided next to it to generate a short link of this URL because it contains all settings and is thus rather long. An advantage of this configuration is that almost all settings can be provided through the URL, which thus can be used as a simple API. This API is not documented yet, but should be self-explanatory using a given example.

For further use of the Solar-MACH results, the user can save the plot by clicking with the right mouse button on it and using the web browser functionality, exactly like in the Jupyter Notebook version mentioned in Sects.~\ref{sect:python_package} and \ref{sect:pfss}. In addition, the web tool provides a download button for the plot that might be helpful, especially when using mobile devices like smartphones or tablets. The web tool also automatically displays a table with all ephemeris, magnetic footpoint, and separation information, which can be downloaded as a comma-separated values (CSV) file.

The Streamlit application itself consists of a rather small code base that is open source \citep{Solar-MACH2022}. It collects all necessary parameters from the user through the GUI, hands them over to the \texttt{solarmach} package described in Sect.~\ref{sect:python_package} , and presents the returned results. The online version provided at \href{https://solar-mach.github.io}{solar-mach.github.io} is automatically redeployed whenever the corresponding \href{https://github.com/jgieseler/Solar-MACH}{GitHub repository} is updated. In this process, the latest \texttt{solarmach} version available at PyPI is installed. Because the GUI needs to be adjusted manually to any new functionality, the Streamlit version may not always offer the latest additions to the \texttt{solarmach} package.


\section{Discussion} 
This paper introduces the Solar-MACH tool to the scientific community. Solar-MACH automatically obtains and visualizes the constellation of the heliospheric spacecraft fleet for a given time and provides a first estimation of the magnetic connectivity of all observers. Here, we illustrated the general workflow of the tool, presented the corresponding Python package \texttt{solarmach} and its PFSS extension, and described the important web version of it provided through the \texttt{streamlit} library. While the tool is intended to have a very simple interface, it is important to point out that for the magnetic backmapping several assumptions are made in the background. Those results should therefore only be interpreted by users with sufficient knowledge of the topic.

The development process of Solar-MACH started at the University of Kiel, Germany, and further discussion took place within the ESA \href{https://www.cosmos.esa.int/web/esdc/archives-user-groups/heliophysics}{Heliophysics Archives USer (HAUS) group}. Now, the main development continues within the SERPENTINE EU Horizon 2020 project, which has been presented in the beginning of this paper. It is important to note that the project is, and remains, fully open source (under the BSD 3-Clause License) and very much encourages everyone to contribute, either by actual code or by any other kind of feedback.

The software is completely developed in Python and uses established open-source Python libraries such as \texttt{pfsspy}, \texttt{streamlit}, and \texttt{sunpy}. The development takes place continuously in public GitHub repositories (separated for the Python package \texttt{solarmach} and the web application), which are also the main entry point for any kind of participation. We encourage users to give feedback on bugs or possible new features by submitting GitHub issues. This should also be the starting point for code contributions. But of course we are also open to feedback through direct communication.
All code versions released on GitHub are automatically archived at Zenodo with a DOI, so that each version (or the latest) can be cited independently.

Solar-MACH is already used widely in the scientific community \citep[e.g.,][]{Badman2022, Hu2022, Mierla2022, Papaioannou2022, Rodriguez-Garcia2022}, but the development is still ongoing with the introduction of additional functionalities. Some important future plans have already been mentioned in this paper, such as fully integrating the PFSS extension into Solar-MACH. Further ideas are listed in the GitHub repository's \href{https://github.com/jgieseler/solarmach/issues}{issues} section and contain, among others, using dynamically obtained solar wind speeds for the HMF lines, adding trajectories of spacecraft, or introducing features related to CME analyses, such as providing a reference cone instead of an arrow. Furthermore, it is planned to incorporate functions that have already been demonstrated in the Jupyter Notebooks into the core \texttt{solarmach} package, for example, the creation of animated GIF files with time-lapses of the spacecraft constellation. For this, like for all other Solar-MACH functions, one important aspect is to provide them in a way such that they remain accessible to all levels of Python users.

\section*{Conflict of Interest Statement}
Author JvF was employed by the company Paradox Cat GmbH since 2021. This article is based on his research conducted prior to that at the University of Kiel and unrelated to his current commercial affiliation. The remaining authors declare that the research was conducted in the absence of any commercial or financial relationships that could be construed as a potential conflict of interest.

\section*{Author Contributions}
ND proposed the initial idea of Solar-MACH and discussed it with VG, AM, MR, and AV.
Based on code by ND, the initial software version was designed by ND and JvF.
JG continued the development.
JG developed the Python package and the Streamlit version.
JG, ND, CP, and RV designed further functionalities.
CP and ND developed the pfsspy integration.
AK, DP, LRG, and DT tested the software and provided critical comments for the development.
JG and RV prepared the first paper draft, and all authors were involved in the preparation of the final manuscript.
All authors revised the manuscript before submission.

\section*{Funding}
This study has received funding from the European Union’s Horizon 2020 research and innovation programme under grant agreement No.\ 101004159 (SERPENTINE).
JG, CP, RV, and DP acknowledge the support of Academy of Finland (FORESAIL, grants 312357, 336809, and 336807).
ND acknowledges the support of Academy of Finland (SHOCKSEE, grant 346902).
JvF thanks the German Space Agency (Deutsches Zentrum für Luft- und Raumfahrt e.V., DLR) for their support of his work on the Solar Orbiter EPD team at the University of Kiel under grant 50OT2002.
A.K. acknowledges financial support from NASA’s NNN06AA01C (SO-SIS Phase-E) contract.
LRG acknowledges the financial support by the Spanish Ministerio de Ciencia, Innovación y Universidades FEDER/MCIU/AEI Projects ESP2017-88436-R and PID2019-104863RB-I00/AEI/10.13039/501100011033.

\section*{Acknowledgments}
The authors acknowledge the Heliophysics Archives USer (HAUS) group at ESA, where the idea for this tool was initially discussed.
The authors would like to thank everyone who helped to improve Solar-MACH by providing feedback or contributing to the various open-source projects that it utilizes.
The authors acknowledge the free cloud hosting provided by Streamlit Inc.
The authors thank the two reviewers for their valuable feedback.


\section*{Data Availability Statement}
The source code used in this study can be found online:
\begin{itemize}
    \item Python package code in GitHub repository \href{https://github.com/jgieseler/solarmach}{solarmach}, archived at \url{https://doi.org/10.5281/zenodo.7311178}
    \item Streamlit code in GitHub repository \href{https://github.com/jgieseler/Solar-MACH}{Solar-MACH}, archived at \url{https://doi.org/10.5281/zenodo.7311215}
    \item Extended example Jupyter Notebook in GitHub repository \href{https://github.com/serpentine-h2020/serpentine}{serpentine}, archived at \url{https://zenodo.org/record/7104866}
\end{itemize}


\def\aj{AJ}%
\def\actaa{Acta Astron.}%
\def\araa{ARA\&A}%
\def\apj{ApJ}%
\def\apjl{ApJ}%
\def\apjs{ApJS}%
\def\ao{Appl.~Opt.}%
\def\apss{Ap\&SS}%
\def\aap{A\&A}%
\def\aapr{A\&A~Rev.}%
\def\aaps{A\&AS}%
\def\azh{AZh}%
\def\baas{BAAS}%
\def\bac{Bull. astr. Inst. Czechosl.}%
\def\caa{Chinese Astron. Astrophys.}%
\def\cjaa{Chinese J. Astron. Astrophys.}%
\def\icarus{Icarus}%
\def\jcap{J. Cosmology Astropart. Phys.}%
\def\jrasc{JRASC}%
\def\mnras{MNRAS}%
\def\memras{MmRAS}%
\def\na{New A}%
\def\nar{New A Rev.}%
\def\pasa{PASA}%
\def\pra{Phys.~Rev.~A}%
\def\prb{Phys.~Rev.~B}%
\def\prc{Phys.~Rev.~C}%
\def\prd{Phys.~Rev.~D}%
\def\pre{Phys.~Rev.~E}%
\def\prl{Phys.~Rev.~Lett.}%
\def\pasp{PASP}%
\def\pasj{PASJ}%
\def\qjras{QJRAS}%
\def\rmxaa{Rev. Mexicana Astron. Astrofis.}%
\def\skytel{S\&T}%
\def\solphys{Sol.~Phys.}%
\def\sovast{Soviet~Ast.}%
\def\ssr{Space~Sci.~Rev.}%
\def\zap{ZAp}%
\def\nat{Nature}%
\def\iaucirc{IAU~Circ.}%
\def\aplett{Astrophys.~Lett.}%
\def\apspr{Astrophys.~Space~Phys.~Res.}%
\def\bain{Bull.~Astron.~Inst.~Netherlands}%
\def\fcp{Fund.~Cosmic~Phys.}%
\def\gca{Geochim.~Cosmochim.~Acta}%
\def\grl{Geophys.~Res.~Lett.}%
\def\jcp{J.~Chem.~Phys.}%
\def\jgr{J.~Geophys.~Res.}%
\def\jqsrt{J.~Quant.~Spec.~Radiat.~Transf.}%
\def\memsai{Mem.~Soc.~Astron.~Italiana}%
\def\nphysa{Nucl.~Phys.~A}%
\def\physrep{Phys.~Rep.}%
\def\physscr{Phys.~Scr}%
\def\planss{Planet.~Space~Sci.}%
\def\procspie{Proc.~SPIE}%

\bibliographystyle{Frontiers-Harvard} 
\bibliography{references}






\end{document}